\documentclass[reprint,aps,prc,twocolumn,superscriptaddress]{revtex4-2}

\usepackage{CJK}
\usepackage{graphicx}
\usepackage{hyperref}
\usepackage{amssymb}
\usepackage{amsmath}
\usepackage[normalem]{ulem}
\usepackage{url}
\usepackage{color}
\usepackage{longtable}

\newcommand{\corref}[1]{\textsuperscript{\textdagger}}  

\begin{document}
\begin{CJK*} {UTF8}{} 

\title{Investigation of the neutron-proton effective mass splitting via heavy ion collisions: Constraints and Implications}


%

\author{Junping Yang}
\affiliation{China Institute of Atomic Energy, P. O. Box 275(18), Beijing 102413, China}
\affiliation{College of Physics and Optoelectronic Engineering, Shenzhen University, Shenzhen 518060, China}  
\author{Meiqi Sun}
\affiliation{China Institute of Atomic Energy, P. O. Box 275(18), Beijing 102413, China} 
\author{Ying Cui}
\affiliation{China Institute of Atomic Energy, P. O. Box 275(18), Beijing 102413, China}
\author{Yangyang Liu}
\affiliation{China Institute of Atomic Energy, P. O. Box 275(18), Beijing 102413, China}
\author{Zhuxia Li}
\affiliation{China Institute of Atomic Energy, P. O. Box 275(18), Beijing 102413, China}
\author{Kai Zhao}
\email{zhaokai@ciae.ac.cn}
\affiliation{China Institute of Atomic Energy, P. O. Box 275(18), Beijing 102413, China} 
\author{Yingxun Zhang}
\email{zhyx@ciae.ac.cn}
\affiliation{China Institute of Atomic Energy, P. O. Box 275(18), Beijing 102413, China}  
\affiliation{Department of Physics and Technology, Guangxi Normal University, Guilin 540101, China}

\date{\today}

\begin{abstract}

The neutron-proton effective mass splitting ($\Delta m^*_{np}$) is investigated through analyses of heavy-ion collisions using the improved quantum molecular dynamics (ImQMD) model with both standard and extended Skyrme interactions. By uncovering the strong correlation between the slope of the neutron-to-proton yield ratio with respect to the kinetic energy (i.e., $S_{n/p} $) and $\Delta m^*_{np}$, 
we reveal that the constraints of the neutron-proton effective mass splitting via heavy ion collisions depend on the kinetic energy region of the emitted nucleons. At low kinetic energies, the data favor $m_n^*>m_p^*$ which is consistent with the nucleon-nucleus scattering analysis, while at high kinetic energies, they favor $m_n^*<m_p^*$. Our findings partly resolve the longstanding discrepancy in the constraints of neutron-proton effective mass splitting with heavy ion collisions and nucleon-nucleus scattering, and significantly advance the understanding of nucleon effective mass splitting through heavy ion collisions. 

\end{abstract}


\maketitle
\end{CJK*}

The nucleon effective mass is utilized to describe the motion of nucleons in a momentum-dependent potential, equivalent to the movement of a quasi-nucleon with mass $m^*_N$ in a momentum-independent potential~\cite{BALi2018PPNP}. In isospin asymmetric nuclear matter, the difference between the effective masses of neutrons and protons ($m^*_n$ and $m^*_p$) is usually named as the isospin splitting of the nucleon effective mass. The neutron-proton effective mass splitting is a crucial microscopic input for studying the thermal properties of proto-neutron stars\cite{Nakazato_2019,Madappa1997,Carolyn2019}, liquid-gas phase transition\cite{JunXu}, the shear viscosity of nuclear matter\cite{XuJun2015PRC}, in-medium nucleon-nucleon cross section\cite{Persram02xs,BALi05xs,YCui18}, and the mechanism of the heavy ion collisions\cite{YXZhang14PLB, Coupland2016, ZhaoqingFeng,JunSu17,JunSu2020,ZhangF2020, FYWang2023NST,WJXie2014PLB}. 
However, the sign and magnitude of the neutron-proton effective mass splitting have not been well known due to the poor knowledge about the isovector channel of the in-medium nucleon-nucleon interaction\cite{XLShang21,WangSB2023PRC,ZuoW2005PRC,Whitehead2021,Mansour2010POTP,Margueron2018,Chamseddine2023}, and the constraints on the neutron-proton effective mass splitting are still in hot debates.

\begin{figure}[htbp]
\centering
\includegraphics[width=\linewidth]{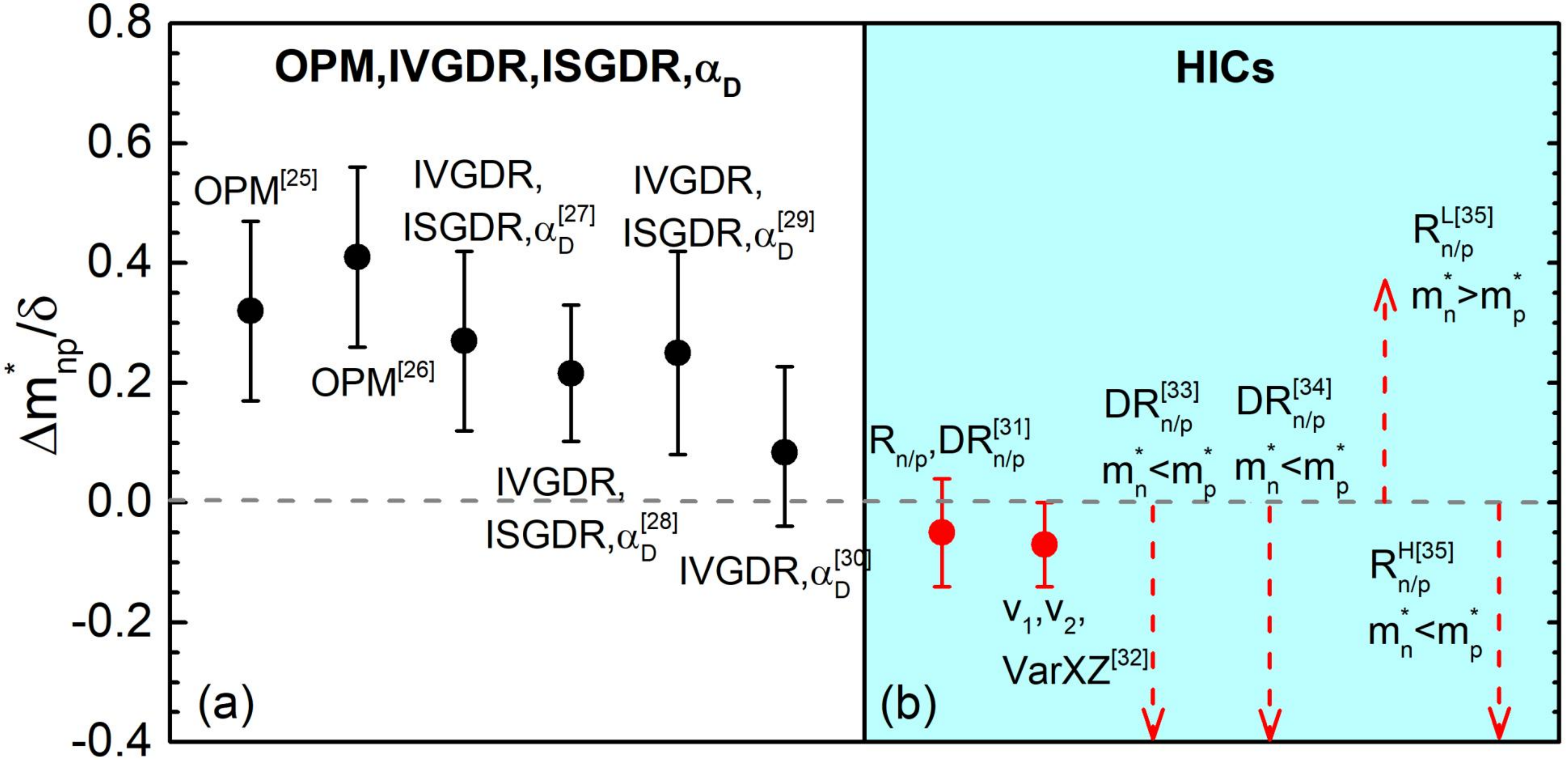}
\setlength{\abovecaptionskip}{0pt}
\vspace{0.2cm}
\caption{The constraints on the neutron-proton effective mass splitting 
by different methods\cite{ChangXu2010PRC,XiaoHuaLi,ZhangZhen2016PRC,KongHY2017,XuJunPRC2020,XuJun2020PLB,Morfouace2019,CYTSang2024,XieWJPRC2013,SuJun2016,YJP2024PRC}.
}
\setlength{\belowcaptionskip}{0pt}
\label{Effe-methods}
\end{figure}

Up to now, the experimental constraints on the neutron-proton effective mass splitting are mainly determined via the following ways, such as the nucleon-nucleus scattering\cite{ChangXu2010PRC,XiaoHuaLi}, isoscalar and isovector giant resonance\cite{ZhangZhen2016PRC,KongHY2017,XuJunPRC2020,XuJun2020PLB}, and heavy ion collisions\cite{Morfouace2019,CYTSang2024,XieWJPRC2013,SuJun2016,YJP2024PRC}. In Figure ~\ref{Effe-methods}, we present part of the results from these experimental efforts. The analyses of nucleon-nucleus scattering data with optical potential models have limited the effective mass splitting at normal density as $\Delta m_{np}^*=\frac{m_n^*-m_p^*}{m}=(0.41\pm0.15)\delta$\cite{XiaoHuaLi}, here $\delta=(\rho_n-\rho_p)/(\rho_n+\rho_p)$ is the isospin asymmetry. For the above analysis, one should note that the results are for normal density and the nucleon kinetic energy in nuclear matter is less than 100 MeV\cite{BALi2018PPNP}. The Bayesian model analysis of isoscalar and isovector giant resonance information in $^{208} \mathrm{~Pb}$ has yielded an effective mass splitting $\Delta m_{n p}^*=$ $\left(0.084_{-0.123}^{+0.143}\right) \delta$\cite{XuJun2020PLB}. For HIC data, the Bayesian analysis of neutron and proton single ratios $R_{n / p}$ and double ratios $D R_{n / p}$, show that the parameter sets with the nucleon effective mass splitting $\Delta m_{n p}^*=(-0.05 \pm 0.09) \delta$\cite{Morfouace2019} can reproduce the data. More recently, a combination analysis by using the S$\pi$RIT experiment measurements of flow and stopping power data along with the ImQMD model was performed and constrained values of the effective mass splitting are $\Delta m_{n p}^*=\left(-0.07_{-0.06}^{+0.07}\right) \delta$\cite{CYTSang2024}, which is consistent with the constraints by using the single neutron to proton yield ratios ($R_{n/p}$) and double neutron to proton yield ratios ($DR(n/p)$) data. Obviously, the constraints on the sign and magnitude of effective mass splitting depend on the methods. On the side of HICs, the possible reasons for this discrepancy could be due to the complex collision mechanism, and the lack of sensitive experimental observables that is strongly correlated with the neutron-proton effective mass splitting. 

\setlength{\abovedisplayskip}{3pt}

In our previous work\cite{YJP2024PRC}, we found that 
the slope of the neutron to proton yield ratios ($R_{n/p}$) with respect to the kinetic energy of emitted nucleons ($E_k$), i.e., $S_{n/p}$, depend on the neutron-proton effective mass splitting as,
\begin{equation}\label{eq:Snp}
    S_{n/p}=\frac{\partial \ln R_{n/p}}{\partial E_k}\propto-(\frac{m}{m_s^*})^2\Delta m_{np}^*,
\end{equation}
where $m_s^*$ is the isoscalar effective mass. 
By comparing the data of $S_{n/p}$ with the ImQMD calculations, the evidence that the emitted nucleons with different kinetic energies favor the parameter sets with different isospin splitting of the nucleon effective mass was firstly discussed. 

However, only the slope of symmetry energy $L$ and the strength of the effective mass splitting $\Delta m_{np}^*$ were varied in Ref.\cite{YJP2024PRC}.  They were taken as $L$=46 MeV and 100 MeV, and $\Delta m_{np}^*=\pm 0.35\delta$. While these variations are insightful, they may not fully account for the influence of other factors, such as the symmetry energy coefficient $S_0$, and isoscalar effective mass $m_s^*/m$, because symmetry energy coefficient $S_0$ and the isoscalar effective mass $m_s^*/m$ were fixed at $S_0=32~\text{MeV}$ and $m_s^*/m=0.77$ in Ref.\cite{YJP2024PRC}. To further understand the constraints of the isospin splitting of the neutron-proton effective mass from HICs and the possible reasons for the discrepancy between the data and the transport model calculations, it is necessary to conduct the analysis in the multi-parameter space, such as the parameter space consisted with $S_0$, $L$,  $m_s^*/m$, and $\Delta m_{np}^*$.

In this work, we calculate the isospin sensitive observable, the single and double neutron to proton yield ratios, i.e., $R_{n/p}$ and $DR(n/p)$, and its slope with respect to kinetic energy for $^{112,124}$Sn+$^{112,124}$Sn in the multi-parameter space by using the ImQMD model\cite{YJP2024PRC}. Then we analyze the sensitivity of $S_{n/p}$ and $S_{DR}$ to the nuclear matter parameters, and obtain the values of $\Delta m_{np}^*$ by comparing the calculations to the data. Finally, the discrepancy between the calculations with Skyrme effective interaction and the data is discussed.

The version of ImQMD model we used is as the same as that in Ref.\cite{YJP2024PRC}, the Skyrme potential energy density without the spin-orbit term is used,
\begin{equation}
\label{eq:edf-imqmd}
    u_\text{sky}=u_\text{loc}+u_\text{nonloc}.
\end{equation}
The local potential energy density is,
\begin{eqnarray} \label{eq:edfimqmd}
\begin{aligned}
    u_\text{loc} &= \frac{\alpha}{2}\frac{\rho^2}{\rho_0} +\frac{\beta}{\gamma+1}\frac{\rho^{\gamma+1}}{\rho_0^\gamma}+\frac{g_\text{sur}}{2\rho_0 }(\nabla \rho)^2 \\
    & \quad +\frac{g_\text{sur,iso}}{\rho_0}[\nabla(\rho_n-\rho_p)]^2 \\
    & \quad +A_\text{sym}\frac{\rho^2}{\rho_0}\delta^2+B_\text{sym}\frac{\rho^{\gamma+1}}{\rho_0^\gamma}\delta^2.
\end{aligned}
\end{eqnarray}
$\rho=\rho_n+\rho_p$ is the nucleon density and $\delta=(\rho_n-\rho_p)/\rho$ is the isospin asymmetry. The $\alpha$ is the parameter related to the two-body term. $\beta$ and $\gamma$ are related to the non-linear density dependent interaction term. $g_\text{sur}$ and $g_\text{sur,iso}$ are related to the surface terms. $A_\text{sym}$ and $B_\text{sym}$ are the coefficients in the symmetry potential and come from the two-body and non-linear density dependent interaction terms\cite{Zhang06,Zhang20FOP}.

The nonlocal potential energy density $u_\text{nonloc}$ in Eq.(\ref{eq:edf-imqmd}) is used as,
\begin{eqnarray} \label{eq:extmd-imqmd}
\begin{aligned}
    u_\text{nonloc} & = \tilde{C}_0\sum_{ij}\int \text{d}^3p\text{d}^3p' f_i(\mathbf{r},\mathbf{p})f_j(\mathbf{r},\mathbf{p'})g(\mathbf{p}-\mathbf{p'}) \\
    & + \tilde{D}_0\sum_{ij\in n}\int \text{d}^3 p \text{d}^3p' f_i(\mathbf{r},\mathbf{p}) f_j(\mathbf{r},\mathbf{p'})g(\mathbf{p}-\mathbf{p'}) \\
    &+ \tilde{D}_0\sum_{ij\in p}\int \text{d}^3p \text{d}^3p' f_i(\mathbf{r},\mathbf{p}) f_j(\mathbf{r},\mathbf{p'})g(\mathbf{p}-\mathbf{p'}).
\end{aligned}
\end{eqnarray}
Here, $g(\mathbf{p}-\mathbf{p'})$ is a phenomenological momentum-dependent interaction (MDI). For extended Skyrme MDI, it is written as,
\begin{eqnarray}
\label{eq:ext-mdi}
\begin{aligned}
g(\mathbf{p}-\mathbf{p'})&=\sum_{I=0}^N b_I (\mathbf{p}-\mathbf{p}')^{2I}. 
\end{aligned}
\end{eqnarray}
The number of $N$, the interaction parameters $b_I$, $\tilde{C}_{0}$ and $ \tilde{D}_{0}$ in Eq.(\ref{eq:ext-mdi}) and Eq.(\ref{eq:extmd-imqmd}) are determined by fitting the Hama optical potential data\cite{hama1990} and the details can be found in Ref.\cite{YJP2024PRC}. The parameter $b_I$ is used to determine the shape of the MDI, and its dimension is GeV$^{2-2I}$ for keeping the dimension of $g(\mathbf{p}-\mathbf{p'})$ in GeV$^2$. For the standard Skyrme MDI, only $(\mathbf{p}-\mathbf{p}')^2$ term is kept and the coefficients are determined from the standard Skyrme interaction sets as in Ref.\cite{YXZhang14PLB}. 

The values of parameters, such as $\alpha$, $\beta$, $\gamma$, $A_\text{sym}$, $B_\text{sym}$, $\tilde{C}_0$ and $\tilde{D}_0$, are related to the nuclear matter parameters such as the saturation density $\rho_0$, binding energy at saturation density $E_{0}$, incompressibility $K_{0}$, isoscalar effective mass $m_s^*$, symmetry energy coefficient $S_{0}$, slope of symmetry energy $L$, and effective mass splitting $\Delta m_{np}^*$, according to the relationship in Ref.\cite{YXZhang20,YJP2024PRC}. 
However, the exact value of the $\Delta m_{np}^*$ depends on the number of the Taylor expansion of isospin asymmetry\cite{BALi2018PPNP,FYWang2023NST}, which means that the different truncations will lead to the different numerical errors.  
To avoid the truncation error in the practical calculations, we use the difference of the inverse of neutron and proton effective mass, i.e., 
\begin{equation}
f_I=\frac{1}{2\delta}\left(\frac{m}{m_n^*}-\frac{m}{m_p^*} \right) =\frac{m}{m_s^*}-\frac{m}{m_v^*}   
\end{equation}
as an input for ImQMD model rather than $\Delta m_{np}^*$\cite{YXZhang14PLB,FYWang2023NST}. The advantage of using the $f_I$ as the input for the transport model is that the values of $f_I$ depend on the isoscalar effective mass and isovector effective mass ($m_v^*$), and do not depend on the isospin asymmetry of the system.

The nuclear matter parameters, such as $\rho_0$, $E_{0}$ and $K_{0}$, and surface coefficients $g_\text{sur}$ and $g_\text{sur,iso}$, are fixed in the calculations, and their values are listed in Table~\ref{tab:MP-Criteria}. The parameters  $S_{0}$, $L$, $m_s^*$ and $f_I$ are variable. 
In our calculations, the prior knowledge on the parameters $m_s^*$ and $f_I$ are obtained according to the current knowledge on the nuclear matter parameters\cite{YXZhang20}, i.e., $m^*_s/m=[0.6,1.0]$, $f_I=[-0.5,0.4]$, and the prior knowledge on the parameters are $S_{0}=[28,41]$ MeV and $L=[30,134]$ MeV are obtained in Refs.\cite{Lynch2018,Burgio2021,Kumar2024,YXZhang20}. 

\setlength{\tabcolsep}{4.8pt}
\begin{table}[htbp]
\caption{\label{tab:MP-Criteria}%
The values of the nuclear matter parameters used in this work. The dimensions of $E_{0}$, $K_{0}$, $S_{0}$, $L$ are in MeV. $m_s^*/m$ and $f_I$ are dimensionless. $g_\text{sur}$=24.56 MeVfm$^2$ and $g_\text{sur,iso}$=-4.99 MeVfm$^2$.}
\centering
\begin{tabular}{ccccccccc}
\hline
\hline
 $ \rho_0 $ & $ E_0 $ & $ K_0 $ & $ S_0 $ & $ L  $ & $ m_s^*/m $ &  $f_I$ 
 \\ \hline
 0.160 & -16 & 230 & [28,41] & [30,134] & [0.6,1.0] & [-0.5,0.4] \\
\hline
\hline
\end{tabular}
\end{table}




To understand the sensitivity of the isospin sensitive observable $S_{n/p}$ to the nuclear matter parameters, such as  $S_0$, $L$, $m_s^*$, and $f_I$, we vary them within the range listed in Table~\ref{tab:MP-Criteria}. 
189 parameter sets are sampled by using Latin hypercube sampling method, 
and the central collisions of the systems $A=^{124}$Sn+$^{124}$Sn and $B=^{112}$Sn+$^{112}$Sn are simulated. The beam energy and impact parameter are set as 120 MeV/u and $b=2fm$. The total number of simulated events for each system is 50,000. 

\begin{figure}[htbp]
\centering
\includegraphics[width=\linewidth]{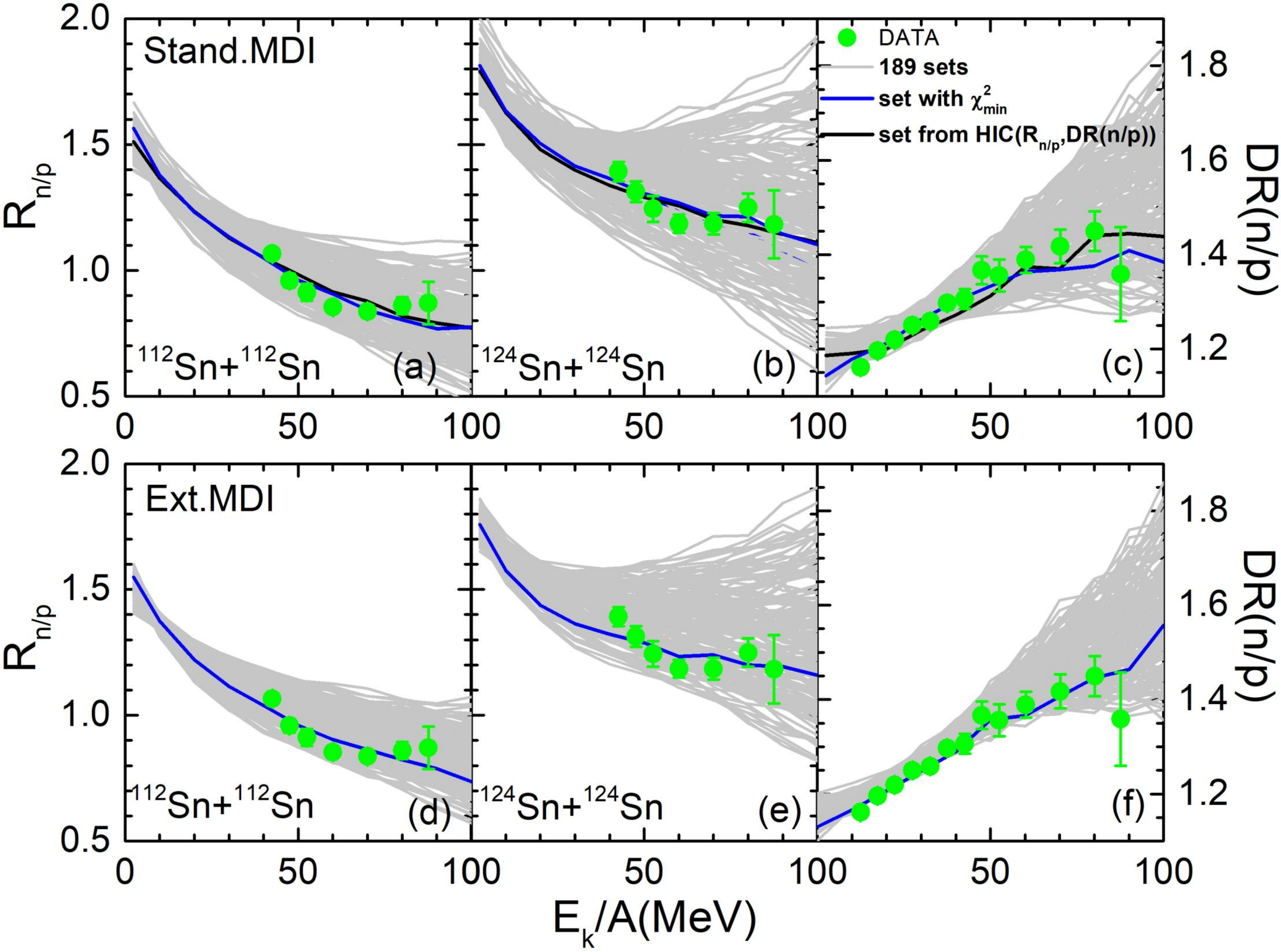}
\setlength{\abovecaptionskip}{0pt}
\vspace{0.2cm}
\caption{The calculated results on the coalescence invariant $R_{n/p}$ and $DR(n/p)$ as functions of $E_k/A$ for $^{112,124}\mathrm{Sn}+^{112,124}\mathrm{Sn}$ at the beam energy $ E_\text{beam} = 120~\text{MeV/u} $. Blue (black) curve is the result obtained with the optimal parameter set in this work (in Ref.\cite{Morfouace2019}), and solid symbols are the experimental results from Ref.\cite{Morfouace2019}.} 
\setlength{\belowcaptionskip}{0pt}
\label{100set-Rnp}
\end{figure}

First, we check the single ratio of the coalescence invariant (CI) neutron and proton, i.e.,  
\begin{equation}
   R_{n/p}=Y_{CI}(n)/Y_{CI}(p), 
\end{equation}
and the double ratio of the coalescence invariant (CI) neutron and proton 
\begin{equation}
   DR(n/p)=R_{n/p}(A)/R_{n/p}(B). 
\end{equation}
The $Y_{CI}(n)$ and $Y_{CI}(p)$ are obtained by combining the free nucleons with those bound in light isotopes with $1< A <5$ as same as in Ref.~\cite{Coupland2016,Morfouace2019}. In Figure~\ref{100set-Rnp} (a)-(f), we present the $R_{n/p}$ and $DR(n/p)$ as a function of the kinetic energy per nucleon of emitted particles in the center-of-mass frame, i.e., $E_k/A$, obtained with the standard (gray lines in top panels) and extended (gray lines in bottom panels) momentum-dependent interactions. 
The green points are the experimental data\cite{Morfouace2019}. 

\setlength{\abovedisplayskip}{3pt}

By performing the $\chi^2$ analysis on the energy spectra of $R^{112}_{n/p}$, $R^{124}_{n/p}$, and $DR(n/p)$ results, one can get the favored parameter sets. 
The blue curves in Figure~\ref{100set-Rnp} are the results obtained with extended and standard Skyrme MDI and the optimal parameter set, i.e., with $\chi^2_{r, min}$. Nonetheless, both of our calculations and that in Ref.\cite{Morfouace2019} show that the curve of $R_{n/p}$ is not exactly reproduced, especially for $S_{n/p}$ as a function of $E_k/A$. 
It means that the constraints should be further understood.

\begin{figure*}[hbpt]
\flushleft
\includegraphics[width=\linewidth]{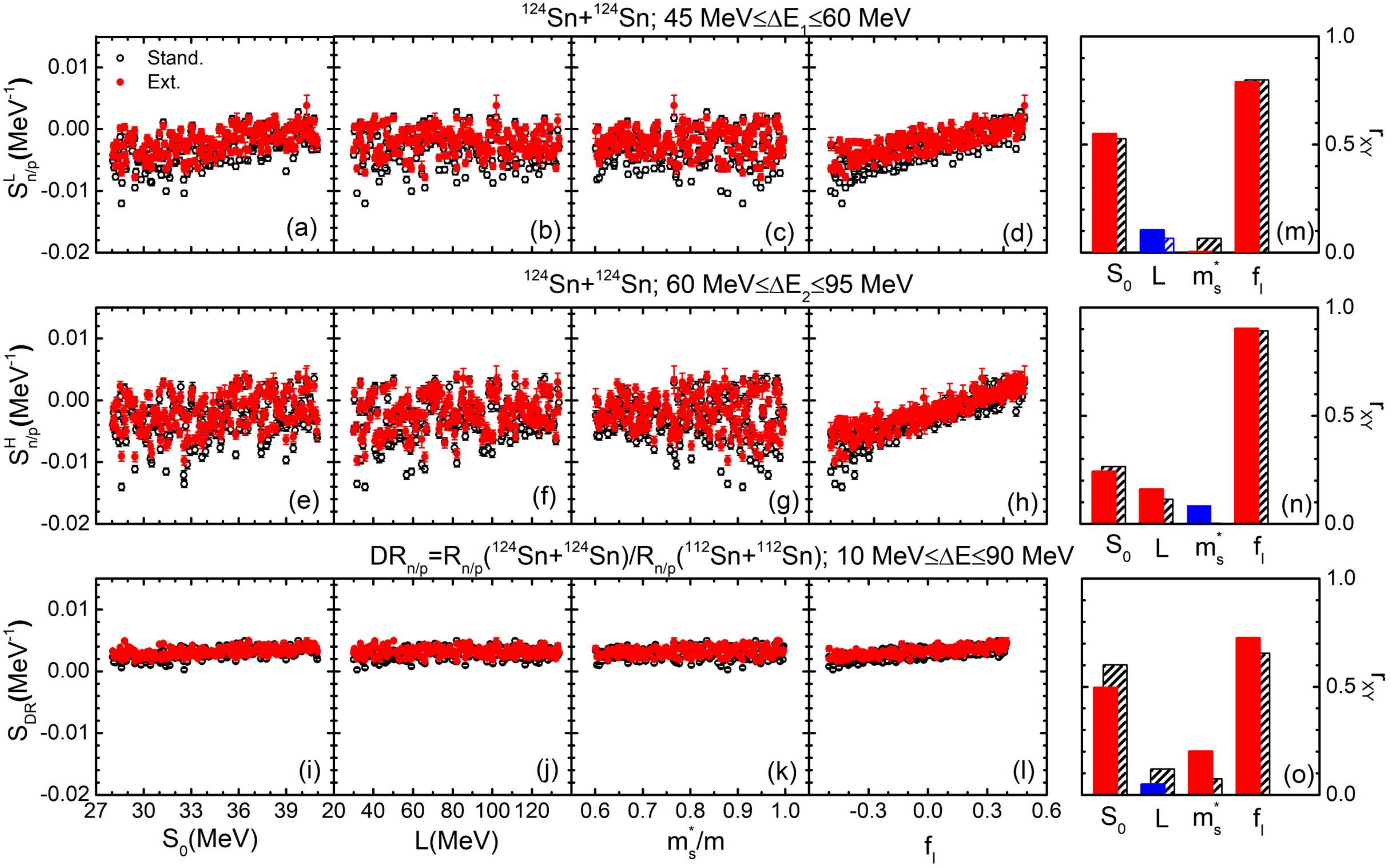}
\setlength{\abovecaptionskip}{0pt}
\vspace{0.2cm}
\caption{(Color online)  Results of $S_{n/p}$ as functions of $S_0$, $L$, $m_s^*$ and $f_I$ for $^{124}\mathrm{Sn}+^{124}\mathrm{Sn}$ obtained with 189 parameter sets at different kinetic regions. Panels (a)-(h) are for $S^{L,H}_{n/p}$, and (i)-(l) are for $S_{DR}$. Red and black symbols are for extended and standard Skyrme MDI, respectively. Panels (m)-(o) are the linear correlation coefficient $r_{XY}$, and the columns with the red (blue) color represent the positive(negative) correlations.}
\setlength{\belowcaptionskip}{0pt}
\label{Snp-mutipara}
\end{figure*}

In the following, the sensitivity of the $S_{n/p}$ to nuclear matter parameters is analyzed. Ideally, one should examine the spectrum of $S_{n/p}$, i.e., $S_{n/p}(E_k/A)$, to accurately constrain the energy dependence of the effective mass splitting. However, the available data of $S_{n/p}$ are limited and subjec to large uncertainties. In our work, we therefore adopt a simply approach, in which the $S_{n/p}$ is calculated as follows,
\begin{equation}\label{eq:snp-cal}
    S_{n/p}=\frac{lnR_{n/p}(E_2)-lnR_{n/p}(E_1)}{E_2-E_1},
\end{equation}
here, $E_1$ and $E_2$ are the selected starting and ending points on the energy spectrum of $R_{n/p}$. 
The $S_{n/p}$ as functions of nuclear matter parameters are presented in Figure~\ref{Snp-mutipara}, in two different kinetic regions. One is the low kinetic energy region, i.e., $E_1=45$ MeV and $E_2=$ 60 MeV, they are presented as $S^{L}_{n/p}$ in panels (a)-(d). Another is the high kinetic energy region, i.e., $E_1=60$ MeV and $E_2=$ 95 MeV, they are presented as $S^{H}_{n/p}$ in panels (e)-(f). The red (black) circles represent the results obtained with the extended (standard) Skyrme MDI in the ImQMD model. Generally, one can find that the $S_{n/p}$ is strongly correlated to the $f_I$, but wealky correlated to $S_0$, $L$, and $m_s^*/m$. 

\setlength{\abovedisplayskip}{3pt}

To quantitatively describe the sensitivity of  $S_{n/p}$ and $S_{DR}$ to these parameters, we calculate the linear correlation coefficient $r_{XY}$ between $X$=$S_{n/p}$ or $S_{DR}$ and the nuclear matter parameters $Y$=$\{S_0, L, m_s^*/m, f_I\}$. The correlation coefficient $r_{XY}$, 
\begin{equation}\label{eq:rXY}
\begin{aligned}
    r_{XY} &= \frac{ \text{cov} ( X, Y ) }{ \sigma ( X ) \sigma ( Y ) }, \\
\end{aligned}
\end{equation}
is calculated as the same as in Ref.\cite{YXZhang15PLB}. Here, $\text{cov} ( X, Y )$ is the covariance between $X$ and $Y$, and $\sigma(X)$ and $\sigma(Y)$ are the standard deviation of $X$ and $Y$, respectively. 
The correlation coefficient $r_{XY}$ between $S_{DR}$ and all nuclear matter parameters are less than 0.7, while the correlation coefficient $r_{XY}$ between $S_{n/p}$ and $f_I$ reach the largest. For $^{124}\mathrm{Sn}+^{124}\mathrm{Sn}$, the $r_{XY}=$0.798 at the low kinetic energy region. While at the high kinetic energy region, and it increases to 0.904. Further calculations with the standard Skyrme MDI also evidence that this strong correlation between $S_{n/p}$ and $f_I$ is not influenced by the form of MDI we used.

In addition, we also analyze the slope of $DR(n/p)$ with respect to the kinetic energy, i.e., $S_{DR}$, which is defined as,
\begin{equation}
         S_{DR}=\frac{ln DR_{n/p}(E_2)-ln DR_{n/p}(E_1)}{E_2-E_1}.
\end{equation}
As shown in Fig.~\ref{100set-Rnp}, the data of $DR$ linearly increases with the kinetic energy except at $E_k/A=90$ MeV. Thus, we choose $E_1=$ 10 MeV and $E_2=$ 90 MeV in the calculations of $S_{DR}$. As shown in Fig.~\ref{Snp-mutipara} (i)-(l), the sensitivity of $S_{DR}$ to the $f_I$ becomes weak. It means that the constrained values of the neutron-proton effective mass splitting by only using the $DR$ ratios will be subject to large uncertainties, and the better choice is to use the $S_{n/p}$.

\begin{figure}[htbp]
\includegraphics[width=\linewidth]{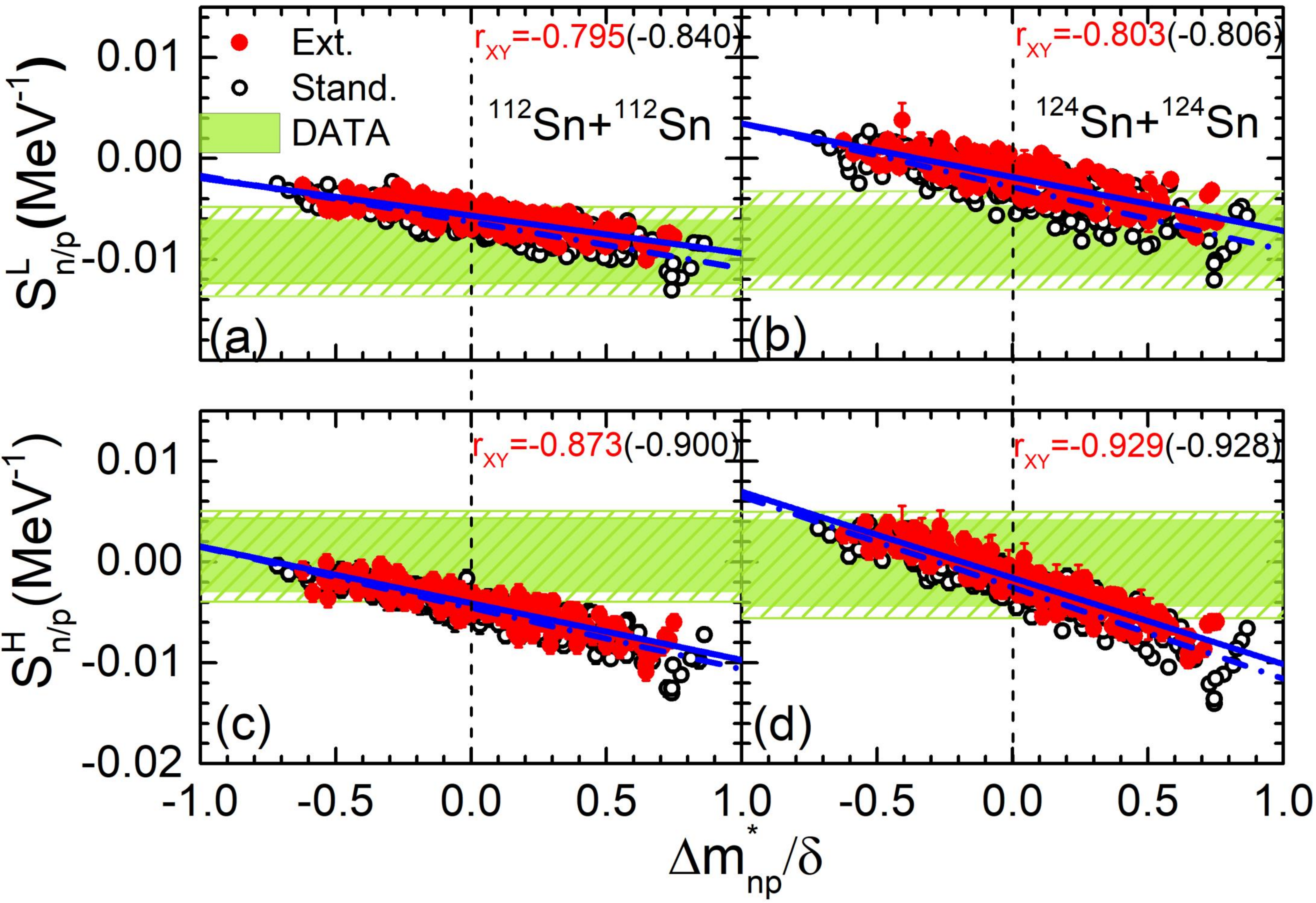}
\caption{(Color online) The $S_{n/p}$ as a function of $\Delta m_{np}^*/\delta$ obtained with 189 parameter sets for the extended MDI (red symbols) and the standard MDI (black symbols). 
The shaded region is the data of $S_{n/p}$ extracted from the published data~\cite{Morfouace2019}. The blue solid and dashed lines are obtained by a linear fitting of the $S_{n/p}$ a function of $\Delta m_{np}^*/\delta$.}
\label{fig:Snp-mnp}
\end{figure}

In Figure~\ref{fig:Snp-mnp},  we replot the $S_{n/p}$ obtained as a function of $\Delta m_{np}^*/\delta$ for two kinetic energy regions. Left panels are the results for $^{112}\mathrm{Sn}+^{112}\mathrm{Sn}$ and right panels are for $^{124}\mathrm{Sn}+^{124}\mathrm{Sn}$. 
The absolute values of $r_{XY}$ reach above 0.80 for two systems at both kinetic energy regions. Especially, the $S^H_{n/p}$ at high kinetic energy for $^{124}\mathrm{Sn}+^{124}\mathrm{Sn}$ is about 0.928, which is larger than the value obtained by using $X=S_{n/p}$ and $Y=f_I$. To guide the eyes on this strong linear correlation relationship, we also plot linear fitting lines in the figure which are presented as blue solid and dashed lines for two kinds of MDIs. Moreover, this conclusion is not affected by the selection of $S_0$ and $L$.

\begin{figure}[htbp]
\includegraphics[width=\linewidth]{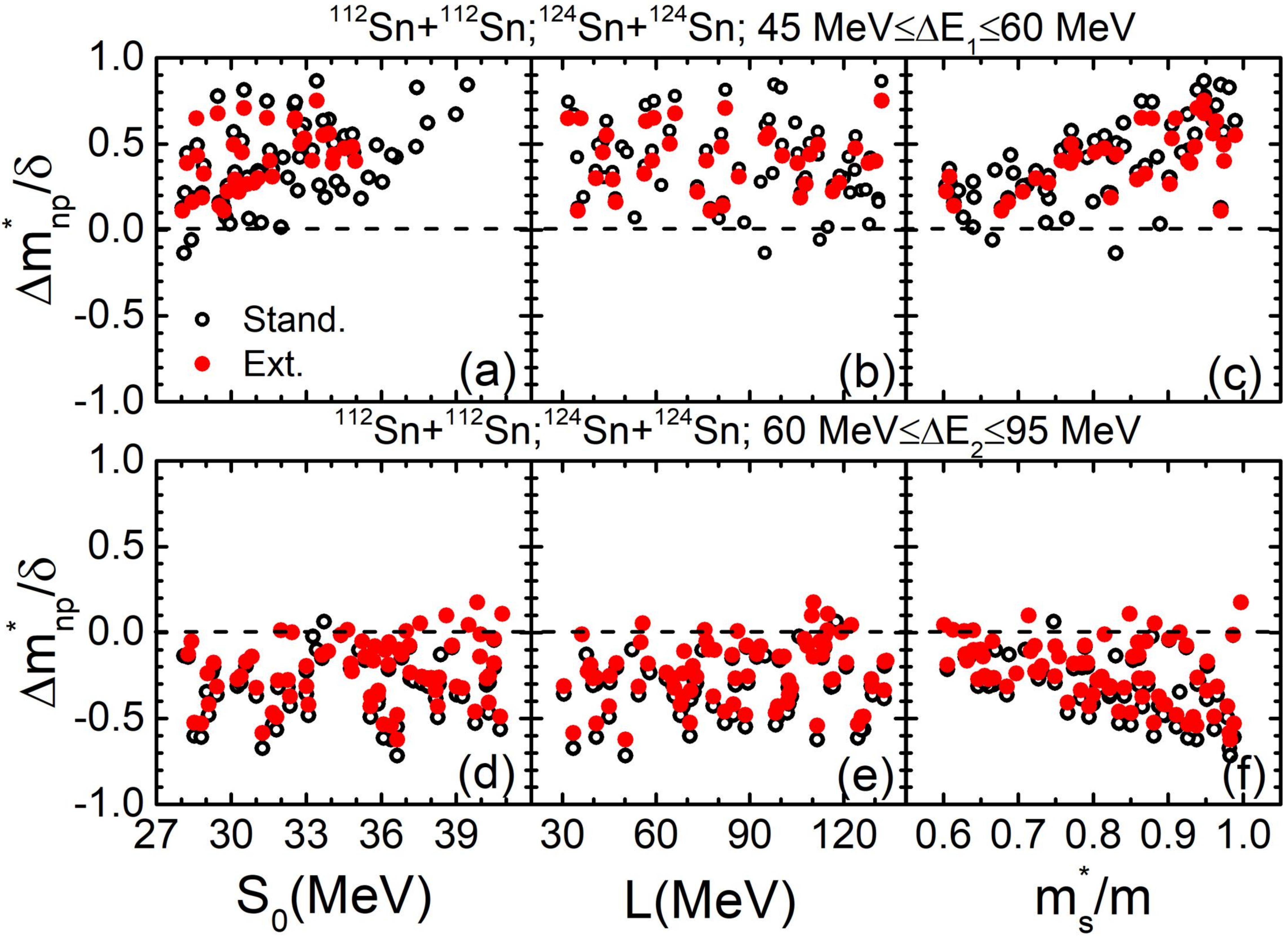}
\caption{(Color online) The constrained values of $\Delta m_{np}^*/\delta$ as a function of $S_0$, $L$, and $m_s^*/m$ by using $S_{n/p}^L$ (upper panels) and $S_{n/p}^H$ (bottom panels), respectively.  The parameter sets that can describe the data are red symbols for the extended MDI and black symbols for the standard MDI.}
\label{fig:mnp-NM-sn112sn124}
\end{figure}

The shaded regions are the $S_{n/p}$ data, which are calculated based on the published data in Ref.~\cite{Morfouace2019}. The associated error is a crucial quantity since it will give the range of the constraints of $\Delta m_{np}^*$. In our work, we calculate the errors with two methods. One is to obtain the error with the difference between the largest and smallest $S_{n/p}$ over the selected kinetic energy region\footnote{The error is $\sigma=\frac{1}{2}|S_{n/p}^{max}-S_{n/p}^{min}|$. The $S_{n/p}^{max}$ and $S_{n/p}^{min}$ are calculated from the upper limit and lower limit of $R_{n/p}$ at the beginning and the end of kinetic energy region, i.e., at $E_1$ and $E_2$.}. This method has been used in Ref.\cite{YJP2024PRC} and is presented as a hatched area. Another method for calculating the error is to use the error propagation formula for Eq.(\ref{eq:snp-cal}), and the result is plotted as a shaded region. 
In the low kinetic energy region, one can find that most of the calculations with the sets $\Delta m_{np}^*/\delta>0$ fall into the data region (with the errors obtained via two methods) for simultaneously describing the data from both systems, as shown in Figure~\ref{fig:mnp-NM-sn112sn124} (a)-(c). Conversely, in the high kinetic energy region, the calculations with the sets $\Delta m_{np}^*/\delta<0$ can fall into the data region (with the errors obtained via two methods), as shown in Figure~\ref{fig:mnp-NM-sn112sn124} (d)-(f).




To quantitatively investigate the constraints on $\Delta m_{np}^*/\delta$ and its correlations with nuclear matter parameters $S_0$, $L$ and $m_s^*/m$, we present the results in Figure~\ref{fig:mnp-NM-sn112sn124} (a)-(f), based on the simultaneously description of the $S_{n/p}^L$ and $S_{n/p}^H$ data. Overall, the constrained values of $\Delta m_{np}^*/\delta$ exhibit a weak dependence on the $S_0$, $L$, and $m_s^*/m$. Nonetheless, a mild correlation between the constrained values of the $m_s^*/m$ and $\Delta m_{np}^*$ is observed. Specifically, when the data of $S_{n/p}^L$ are employed, $\Delta m_{np}^*$ shows a positive correlation with $m_s^*/m$, whereas a negative correlation is found when the $S_{n/p}^H$ data are used. Furthermore, the average values of $\Delta m_{np}^*/\delta$ extracted using $S_{n/p}^L$ range from 0.28 to 0.41, depending on whether the standard Skyrme MDI or extended Skyrme MDI is adopted, and on the choice of error analysis method. In contrast, using the data of $S_{n/p}^H$ yields the average values of $\Delta m_{np}^*/\delta$ ranging from -0.33 to -0.12. The constrained values of $\Delta m_{np}^*$ and other nuclear matter parameters are listed in Table.~\ref{tab:AV-NM-err}. This analysis of $S_{n/p}$ across different energy regimes reveals more detailed insights than previous studies~\cite{Morfouace2019}, which primarily relied on the gross behavior of $R_{n/p}$ and $DR(n/p)$ ratios.



\begin{table*}[htbp]
\caption{\label{tab:AV-NM-err}%
The average value of nuclear matter parameter sets that can simultaneously describe low-energy or high-energy regions under different MDI for two systems. The data of $S_{n/p}$ used in this comparison is extracted from the published data~\cite{Morfouace2019}, and the error is obtained by the error propagation formula (the difference between $S_{n/p}^{max}$ and $S_{n/p}^{min}$).}
\centering
\begin{tabular}{ccccc}
\hline
\hline
 Para. & $D_{Ext}^L$ & $D_{Stand}^L$ & $D_{Ext}^H$ & $D_{Stand}^H$\\
\hline
$S_0$ & 31.27$\pm$2.08(32.37$\pm$3.01) & 32.30$\pm$2.87(33.08$\pm$3.34) & 35.33$\pm$3.64(34.79$\pm$3.70) & 34.81$\pm$3.74(34.92$\pm$3.76) \\
$L$ & 80.68$\pm$30.45(88.52$\pm$31.59) & 85.65$\pm$31.57(86.22$\pm$31.84) & 86.73$\pm$29.06(86.72$\pm$29.00) & 83.98$\pm$29.23(85.94$\pm$29.64) \\
$m_s^*/m$ & 0.84$\pm$0.11(0.81$\pm$0.12) & 0.80$\pm$0.11(0.78$\pm$0.12) & 0.80$\pm$0.12(0.78$\pm$0.12) & 0.83$\pm$0.11(0.80$\pm$0.11) \\
$m_v^*/m$ & 0.66$\pm$0.09(0.67$\pm$0.12) & 0.65$\pm$0.09(0.68$\pm$0.12) & 0.98$\pm$0.23(0.88$\pm$0.24) & 1.04$\pm$0.21(0.94$\pm$0.23) \\ 
$\Delta m_{np}^*/\delta$ & 0.41$\pm$0.18(0.31$\pm$0.22) & 0.39$\pm$0.24(0.28$\pm$0.27) & -0.23$\pm$0.18(-0.12$\pm$0.22) & -0.33$\pm$0.18(-0.21$\pm$0.23) \\
\hline
\hline
\end{tabular}
\end{table*}

The present results for $\Delta m^*_{np}/\delta$ underscore the complexities involved in constraining effective mass splitting via heavy-ion collisions (HICs). In HICs, nucleons emitted with low and high kinetic energies originate from different stages and density regions. Specifically, low-energy nucleons are primarily emitted from the low-density regions during the late stages, while high-energy nucleons predominantly originate from the high-density regions in the early stages. Consequently, nucleons with varying kinetic energies carry information about the symmetry potential at different densities and momenta. This intricate behavior cannot be fully encapsulated by a single Skyrme parameter set characterized by a fixed neutron-proton effective mass splitting ($\Delta m^*_{np}$). The current calculations suggest two key points. First, the symmetry potential may exhibit a non-monotonic dependence on momentum, decreasing initially and then increasing, a feature not achievable with standard Skyrme interactions. Notably, this behavior has been predicted by finite-range Gogny forces, such as D1S and D250 ~\cite{Roshan2014,RChen12,MQSun2025}, which warrants further investigation. Second, the high-momentum tail of the nucleon momentum distribution in the initial nucleus may influence the neutron-to-proton yield ratio ($R_{n/p}$) at high kinetic energies, thereby affecting the interpretation of the experimental data. Understanding both of them not only needs improvements in the mean field potential and short-range correlations in the transport models, but also needs to cross-check with the nucleonic collective flow observables, such as the collective flow of light particles\cite{CYTSang2024} and neutrons, together with higher-precision measurements on $R_{n/p}$ spectra for the reaction systems at different beam energies in the future.

In summary, we simulate the $^{112,124}\mathrm{Sn}+^{112,124}\mathrm{Sn}$ with the new version of the improved quantum molecular dynamics model (ImQMD), in which the standard and extended Skyrme interactions are adopted. 189 parameter sets which are sampled in four dimensional nuclear matter parameter space, i.e., $S_0$, $L$, $m_s^*$, and $f_I$, under the prior information of symmetry energy obtained in Refs.\cite{Lynch2018,Burgio2021,Kumar2024}. Our calculations show that the slope of the emitted neutron to proton yield ratios with respect to the kinetic energy of the emitted nucleons, i.e., $S_{n/p}$, is strongly correlated to the neutron-proton effective mass splitting. The linear correlation coefficient between $S_{n/p}$ and $\Delta m_{np}^*$ reaches above 0.928 for neutron-rich system $^{124}\mathrm{Sn}+^{124}\mathrm{Sn}$.

By comparing the calculated results of $S_{n/p}$ to the data, we find that the effective Skyrme interaction with a fixed neutron-proton effective mass splitting fails to accurately describe the $S_{n/p}$ over the whole kinetic energy region. At the low kinetic energy region (less than 60 MeV), the data favor the neutron effective mass greater than proton effective mass, i.e., $m_n^*>m_p^*$, while the data at high kinetic energy (above 60 MeV) favor the parameter sets with the neutron effective mass less than the proton effective mass, i.e., $m_n^*<m_p^*$. This finding implies that the momentum-dependent symmetry potential could first decrease and then increase with the momentum, which has been predicted in the finite range Gogny interaction\cite{RChen12,Roshan2014,MQSun2025} and the-nucleon collision mechanism\cite{RWada2017}, and it should be investigated via HICs in the future. In another, the effects of high-momentum tail in the initial nucleus\cite{OHen17,ZXYang19} could also influence the energy spectra of the emitted nucleons and it should also be investigated.


\section*{Acknowledgments}

The authors thank the helpful discussions with Profs. Lie-Wen Chen and Xiao-Hua Li. This work was partly inspired by the transport model evaluation project, and it was supported by the National Natural Science Foundation of China under Grants No. 12275359, No. 12375129, No. 11875323 and No. 11961141003, by the National Key R\&D Program of China under Grant No. 2023 YFA1606402, by the Continuous Basic Scientific Research Project, by funding of the China Institute of Atomic Energy under Grant No. YZ222407001301, No. YZ232604001601, and by the Leading Innovation Project of the CNNC under Grants No. LC192209000701 and No. LC202309000201. We acknowledge support by the computing server SCATP in China Institute of Atomic Energy.



\bibliography{References}


\end{document}